\documentclass[aps, pra, notitlepage, superscriptaddress, twocolumn, 10pt, floatfix]{revtex4-2}
\usepackage[colorlinks=true,citecolor=citecolor,linkcolor=citecolor,urlcolor=citecolor]{hyperref}
\usepackage{amsmath}
\usepackage[caption = false]{subfig}
\usepackage{graphicx,epstopdf}
\usepackage[english]{babel}
\usepackage{blindtext}
\usepackage{lipsum}
\usepackage{amsfonts}
\usepackage{amssymb}
\usepackage{bbm}
\usepackage{enumerate}
\usepackage{color}
\usepackage{latexsym}
\usepackage{times,txfonts}
\usepackage{physics}
\usepackage[svgnames]{xcolor}

\definecolor{citecolor}{HTML}{2e3092}

\newcommand{\us}{\uparrow}
\newcommand{\ds}{\downarrow}
\newcommand{\dket}[1]{| #1 \rangle\rangle}

\DeclareSymbolFont{CMletters}{OML}{cmm}{m}{it}
\DeclareMathSymbol{J}{\mathalpha}{CMletters}{`J}
\DeclareMathSymbol{j}{\mathalpha}{CMletters}{`j}
\DeclareMathSymbol{U}{\mathalpha}{CMletters}{`U}

\begin{document} 

\title{Entanglement-enhanced quantum rectification}

\author{Kasper Poulsen}
\email{poulsen@phys.au.dk}
\affiliation{Department of Physics and Astronomy, Aarhus University, Ny Munkegade 120, 8000 Aarhus C, Denmark}

\author{Alan C. Santos}
\email{ac\_santos@df.ufscar.br}
\affiliation{Departamento de F\'{i}sica, Universidade Federal de S\~ao Carlos, Rodovia Washington Lu\'{i}s, km 235 - SP-310, 13565-905 S\~ao Carlos, SP, Brazil}
\affiliation{Department of Physics, Stockholm University, AlbaNova University Center, 106 91 Stockholm, Sweden}

\author{Lasse B. Kristensen}
\affiliation{Department of Physics and Astronomy, Aarhus University, Ny munkegade 120, 8000 Aarhus C, Denmark}

\author{Nikolaj T. Zinner}
\email{zinner@phys.au.dk}
\affiliation{Department of Physics and Astronomy, Aarhus University, Ny munkegade 120, 8000 Aarhus C, Denmark}
\affiliation{Aarhus Institute of Advanced Studies, Aarhus University, Høegh-Guldbergs Gade 6B, 8000 Aarhus C, Denmark}

\begin{abstract}

Quantum mechanics dictates the band-structure of materials that is essential for functional electronic components. 
With increased miniaturization of devices, it becomes possible to exploit the full potential of quantum mechanics through the principles of superposition and entanglement. 
We propose a new class of quantum rectifiers that can leverage entanglement to dramatically increase performance by coupling two small spin chains through an effective double-slit interface. 
Simulations show that rectification is enhanced by several orders of magnitude even in small systems, and that the effect survives in a noisy environment.
Realizable using several of the quantum technology platforms currently available, our findings reveal the importance of quantum entanglement in seemingly contradictory applications such as heat and noise control.
\end{abstract}

\maketitle

\begin{figure*}[t]
\begin{center}
\includegraphics[width=1.0 \linewidth, angle=0]{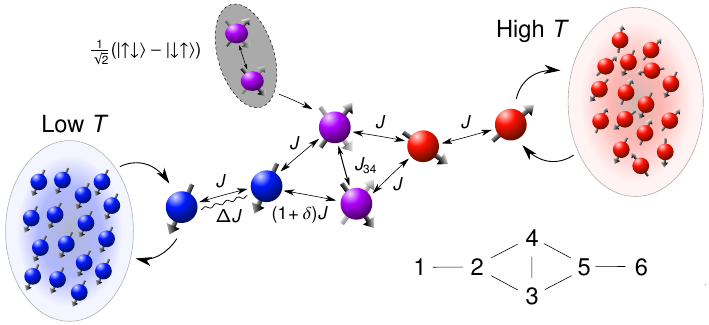}
\end{center}
\caption{Illustration of a few-spin model of a quantum rectification device consisting of two 
segments, $XXZ$ chain on the left and $XX$ chain on the right, connected by a 'two-way' interface. The 
device is connected to thermal baths at each end, one at low and one at high temperature. 
The exchange coupling is $J$, while the two spins in the interface are coupled with an exchange coupling $J_{34}$.
The $Z$ coupling (anisotropy) is $\Delta\, J$ and controls the left-right asymmetry of the diode. 
The dimensionless $\delta$ measures the up-down symmetry-breaking in the system. During operation in reverse bias as shown here, the 
central interface spins are in the maximally-entangled Bell state illustrated in the top left-hand corner.
In the bottom right-hand corner, the numbering of the spins is shown.}
\label{figure1}
\end{figure*}

\section{Introduction}

\noindent Classical electronic components such as transistors and diodes are based on the band-stucture of materials\cite{ashcroft1976solid}, and their integration into circuits and chips constitutes the first quantum revolution. Presently, increased miniaturization requires us to deal with the quantum nature of the information carriers themselves. This is particularly important, as we push towards the new paradigm of quantum computing, and a new toolbox of quantum components needs to be developed.

Transport properties are one of the most basic and essential features of versatile components, and it is hoped that not only charge but also magnetic (spin) \cite{zutic2004,wolf2001, PhysRevLett.126.077203} and  
thermal (phonon) \cite{giazotto2006,ROBERTS2011648,RevModPhys.84.1045,benenti2017} currents can be leveraged in future technologies.
A key component is a current rectifier, well known in electronics as the diode, which features an asymmetry in its forward and reverse transport ability. Schottky or p-n junction diodes are common commercially available designs based on semiconductor materials. However, in recent years, the interest for rectification has spread to other fields such as molecular junctions where molecular diodes have achieved competitive rectification factors \cite{Chen2017}.
Furthermore, important steps towards acoustic \cite{liang2009,liang2010,fleury2014,nomura2019}
and thermal diodes \cite{PhysRevLett.88.094302,PhysRevLett.93.184301,doi:10.1063/1.2191730,chang2006,martinez2015,wang2017}
have been reported recently. 

A particularly promising platform for rectification is quantum spin chains coupled to thermal baths \cite{0295-5075-85-3-37001,PhysRevLett.106.217206,PhysRevLett.110.047201}. Here, a number of spins or two-level systems are connected though XXZ-couplings as well as to thermal baths or magnetic reservoirs. These setups are particularly versatile, and the spin or heat transport can realize components like minimal motors \cite{PhysRevE.95.062143}, thermal transistors \cite{PhysRevLett.116.200601}, thermal 
diodes \cite{PhysRevB.79.014207,PhysRevB.80.172301,PhysRevE.92.062120,PhysRevE.99.032136}, and spin current diodes \cite{PhysRevE.90.042142,loss2011,PhysRevLett.120.200603}. Rectification effects have been found in systems ranging from only one anharmonic molecule \cite{PhysRevB.73.205415} and a system of two spins \cite{PhysRevE.89.062109} to larger 2D geometries \cite{PhysRevE.103.032108} and linear chains in the thermodynamic limit \cite{PhysRevLett.120.200603}.
A common mechanism of these rectifiers is a mismatch of energetics in the vibrational spectra \cite{RevModPhys.84.1045},
or in the electronic \cite{ashcroft1976solid} or the magnetic (spin excitation) 
band gap \cite{PhysRevE.90.042142}.

The theoretical interest is spurred by the increased ability to experimentally study the interplay between quantum degrees of freedom and thermal reservoirs. This includes experimental studies of Maxwell's demon \cite{PhysRevLett.113.030601, PhysRevResearch.2.032025, PhysRevLett.121.030604}, heat engines \cite{PhysRevLett.123.240601, PhysRevLett.125.166802, Josefsson2018}, and heat rectification \cite{Senior2020}. Here, thermal baths or reservoirs are either simulated through stochastic coherent interaction \cite{PhysRevResearch.2.032025} or used directly through e.g. resistors for superconducting circuit platforms \cite{Senior2020} or ferromagnetic leads for spin systems \cite{doi:10.1126/science.1201725}.

Here, we introduce a new class of rectifiers that utilize the quintessential quantum mechanical property of entanglement. By coupling two segments of a quantum spin chain through a two-way junction that entangles the interface spins, we demonstrate boosts of spin and thermal current rectification factors of at least three orders of magnitude even for few-spin systems. The mechanism behind the large spin rectification can be broken into two parts. First, the interface spins become entangled in only one bias. Second, this entanglement blocks transport due to almost perfect destructive interference. Furthermore, we show that the proposed diode has many variations. The effect is seen for a wide range of parameters and in noisy environments, and therefore, it should be realizable using several of the current quantum technology platforms.

\section{Model and results}

To illustrate the mechanism, we concentrate on the few-spin example shown in Fig.~\ref{figure1}. 
It consists of six spin-1/2 particles in a two-segment chain connected by a 'double-slit' interface
and described by an $XXZ$ Heisenberg Hamiltonian of the form
\begin{equation}
\label{hamiltonian_main}
\hat{H}/J = \hat{X}_{12} + (1+\delta) \hat{X}_{23} + \hat{X}_{24} + J_{34}/J \hat{X}_{34} + \hat{X}_{35} + \hat{X}_{45} + \hat{X}_{56} + \Delta \hat{Z}_{12}
\end{equation}
where $\hat{X}_{ij} = \hat{\sigma}_x^{(i)} \hat{\sigma}_x^{(j)} + \hat{\sigma}_y^{(i)} \hat{\sigma}_y^{(j)}$ 
is the $XX$ spin exchange operator, while 
$\hat{Z}_{ij} = \hat{\sigma}_z^{(i)} \hat{\sigma}_z^{(j)}$ is the $Z$ coupling that induces relative
energy shifts. The Pauli matrices for the $i$th spin are denoted $\hat{\sigma}_\alpha^{(i)}$ for $\alpha = x,y,z$, and
we are using units where $\hbar=k_B =1$.  The exchange coupling $J$ gives the overall scale of the problem, while the 
exchange between the interface spins is $J_{34}$. A prerequisite of rectification 
is a breaking of left-right symmetry which we implement by a non-zero $Z$ coupling parametrized
by $\Delta$, although we note that this may as well have been provided by local magnetic fields 
applied to spins 1 and 2 \cite{PhysRevB.79.014207,PhysRevB.80.172301}. 
Due to the interface, we also have to consider up-down symmetry, i.e. the symmetry between the upper and lower part, and we parametrize
its breaking by adding $\delta$ to the exchange between spins 2 and 3 in Fig. \ref{figure1}.
To study rectification of currents in the system, we couple it locally to thermal baths on the left and 
right, see Fig. \ref{figure1}. One bath is cold and forces the adjacent
spin to point down, while the other is hot and forces the adjacent spin into a 
statistical mixture of up and down. The presence of the baths means we have an open 
(non-unitary) quantum system that we describe using the density operator $\hat{\rho}$ and the 
corresponding Lindblad master equation formalism. 
The evolution of the density operator $\hat{\rho}$ of the system is determined by the Lindblad master equation \cite{Lindblad1976,breuer2002theory}
\begin{align}
\frac{\partial \hat{\rho}}{\partial t} = \mathcal{L}[\hat{\rho}] = -i [\hat{H}, \hat{\rho}] + \mathcal{D}_1 [\hat{\rho}] + \mathcal{D}_6 [\hat{\rho}],
\label{Lindblad}
\end{align}
where $[\bullet, \bullet]$ is the commutator, $\mathcal{L}[\hat{\rho}]$ is the Lindblad superoperator, and $\mathcal{D}_n[\hat{\rho}]$ is a dissipative term describing the action of the baths:
\begin{equation}
\begin{aligned}
\mathcal{D}_n [\hat{\rho}] &= \gamma \left[ \lambda_n  \left( \hat{\sigma}_+^{(n)} \hat{\rho} \hat{\sigma}_-^{(n)}  - \frac{1}{2} \left\{ \hat{\sigma}_-^{(n)}\hat{\sigma}_+^{(n)} , \hat{\rho} \right\} \right) \right. \\ &\hspace{1.3cm} \left. +\, (1- \lambda_n) \left( \hat{\sigma}_-^{(n)} \hat{\rho} \hat{\sigma}_+^{(n)}  - \frac{1}{2} \left\{ \hat{\sigma}_+^{(n)}\hat{\sigma}_-^{(n)} , \hat{\rho} \right\} \right) \right],
\end{aligned}
\end{equation}
where $\hat{\sigma}_+^{(n)} = (\hat{\sigma}_-^{(n)})^\dag = (\hat{\sigma}_x^{(n)} + i \hat{\sigma}_y^{(n)}) /2$ and $\{\bullet , \bullet \}$ denotes the anti-commutator. $\gamma$ is the strength of the interaction with the baths which we have set to $\gamma=J$ unless otherwise stated. The nature of the interaction is determined by $\lambda_n$, and we have focused on $\lambda_1$ and $\lambda_6$ set to either 0 or 0.5. If $\lambda_n = 0$, the bath will force the spin to tend down ($\ket{\ds}_n \! \bra{\ds}$) corresponding to a low temperature bath, and if $\lambda_n = 0.5$, the bath will force the spin into a statistical mixture of up and down ($(\ket{\ds}_n \! \bra{\ds} + \ket{\us}_n \! \bra{\us}) /2$) corresponding to a high temperature bath. 
The baths induce currents, and the system is generally in a non-equilibrium state. However, after
sufficient time, it will reach a steady-state (ss), $\dot{\hat{\rho}}_{\mathrm{ss}} = 0$. 
It is this steady-state that determines the rectification properties. 
For $\delta\neq 0$, the steady-state will be unique and independent of the initial state (see Appendix \hyperref[AppA]{A} for further details).
We define the steady-state spin current \cite{PhysRevB.79.014207}
$\mathcal{J} = \mathrm{tr} \{\hat{j}_{12}\hat{\rho}_\mathrm{ss}\}$ as the 
expectation value of the operator
$\hat{j}_{ij} = 2J \left( \hat{\sigma}_{x}^{(i)}\hat{\sigma}_{y}^{(j)} - \hat{\sigma}_{y}^{(i)}\hat{\sigma}_{x}^{(j)}\right)$
in the steady-state. Since the Hamiltonian is spin conserving, the current can be calculated in several ways e.g. $\mathcal{J} = \mathrm{tr} \{\hat{j}_{56}\hat{\rho}_\mathrm{ss}\}$ or $\mathcal{J} = \mathrm{tr} \{[\hat{j}_{23} + \hat{j}_{24}]\hat{\rho}_\mathrm{ss}\}$
By {\it forward bias}, we denote the situation where the hot bath interacts with spin 1, while 
the cold bath interacts with spin 6, and a current $\mathcal{J}_\mathrm{f}$ flows from left to right. In {\it reverse bias}, the cold bath is at spin 1 and the hot bath 
at spin 6 with a (generally negative) current $\mathcal{J}_\mathrm{r}$ flowing from right to left, see Fig. \ref{figure1}.
To obtain a well-functioning diode, we must demand that
\begin{enumerate}
\item no spin current is allowed to flow in reverse bias $\mathcal{J}_\mathrm{r} \sim 0$,
\item an appreciable spin current can flow in forward bias $\mathcal{J}_\mathrm{f} \gg -\mathcal{J}_\mathrm{r}$.
\end{enumerate}
A measure of quality that contains both requirements is the rectification
\begin{equation}
\mathcal{R} = - \frac{\mathcal{J}_\mathrm{f}}{\mathcal{J}_\mathrm{r}},
\end{equation}
which tends to $\mathcal{R} = 1$ when transport is symmetric, while 
a good diode yields $\mathcal{R}\gg 1$. An alternative quality measure
is the contrast defined as
\begin{equation}
\mathcal{C} = \left| 
\frac{\mathcal{J}_\mathrm{f} + \mathcal{J}_\mathrm{r}}{\mathcal{J}_\mathrm{f} - \mathcal{J}_\mathrm{r}} \right|,
\end{equation}
such that $\mathcal{C} = 0$ is equivalent to $\mathcal{R} = 1$, while 
$\mathcal{C} = 1$ for $\mathcal{J}_\mathrm{r}\to 0$, i.e. for the perfect diode.

\begin{figure}[]
\begin{center}
\includegraphics[width=1 \linewidth, angle=0]{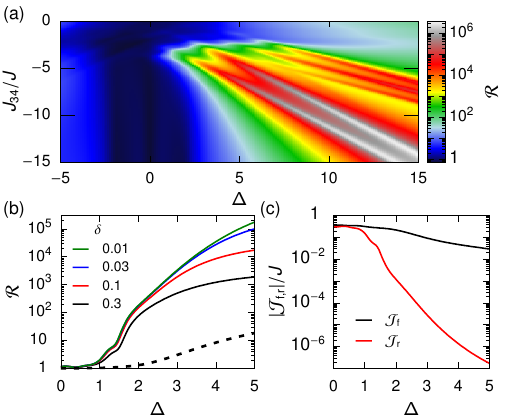}
\end{center}

\caption{(a) $\mathcal{R}$ as a function of $\Delta$ and $J_{34}$ for $\delta = 0.01$. 
(b) $\mathcal{R}$ as a function of $\Delta$ for different values of $\delta$ and $J_{34} = J^{\textrm{c}}_{34}(\Delta)$ (solid lines). 
The dashed line displays $\mathcal{R}$ for a linear chain with spin 3 removed.
(c) Steady-state currents $\mathcal{J}_\mathrm{f}$ and $\mathcal{J}_\mathrm{r}$ for $\delta = 0.01$ and $J_{34} = J^{\textrm{c}}_{34}(\Delta)$. The keys obey the vertical ordering of the graphs.
}
\label{figure2}
\end{figure}

The rectification results for the six-spin implementation of Fig.~\ref{figure1} 
are shown in 
Fig.~\ref{figure2} as a function of the relevant parameters of 
the model. The contour plot in Fig.~\ref{figure2}{(a)} shows 
$\mathcal{R}$ for a small up-down symmetry-breaking of 
$\delta=0.01$ as a function of $J_{34}$ and $\Delta$. Our key 
discovery is the region in the bottom right-hand corner where
values of $\mathcal{R} > 10^6$ are reached. Further inspection of the two lines of large $\mathcal{R}$ shows that they occur for $J_{34} = -(\Delta \pm 1)J$ for large $\Delta$; This precise number will be justified later.  However, very large anisotropi values can be experimentally challenging. Therefore, to keep the model general, we keep $\Delta \leq 5$ for which a better parametrization is $J_{34} = J^{\textrm{c}}_{34}(\Delta)$ where 
\begin{equation}
J^{\textrm{c}}_{34}(\Delta) = -(\Delta + 1.3)J.
\end{equation}
Fig.~\ref{figure2}{(b)} demonstrates the 
dependence on $\Delta$ using $J_{34} = J^{\textrm{c}}_{34}(\Delta)$ showing that $\delta\ll 1$
gives higher $\mathcal{R}$ as a function of $\Delta$. 
This may be advantageous for 
experimental realization as a small asymmetry in the up-down 
symmetry is likely to occur and is a useful control parameter. 
We also confirm that large rectifications
are mainly due to suppression of $\mathcal{J}_\mathrm{r}$, see Fig.~\ref{figure2}{(c)}.

In previous studies \cite{PhysRevB.79.014207,PhysRevLett.120.200603}, it has been shown that significant rectification 
can occur in linear two-segment chains as a function of $\Delta$ due to the band gap induced by the 
$Z$ coupling in one segment. For comparison, the dashed line in Fig.~\ref{figure2}{(b)} shows the rectification when spin 3 in  
Fig.~\ref{figure1} is removed, effectively yielding a linear chain. The increase in rectification, and hence diode quality, of 
the two-way design that includes spin 3 is seen to be three orders of magnitude or more, and is one of our main findings. Removing spin 3 also removes part of the left-right asymmetry through $\delta$. However, removing spin 4 instead results in a linear chain with rectification factors within 1\% of the dashed line in Fig.~\ref{figure2}{(b)}. Therefore, $\delta$ does not contribute significantly for the linear chain.

\subsection{Understanding the mechanism}
\begin{figure}[]
\begin{center}
\includegraphics[width=1 \linewidth, angle=0]{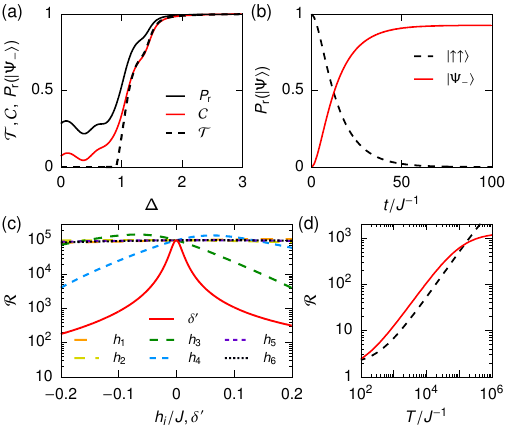}
\end{center}

\caption{(a) Population $P_r( |\Psi_- \rangle )$ (top), Contrast $\mathcal{C}$ (middle), and concurrence $\mathcal{T}$ (bottom) as a function of 
$\Delta$ where $\delta = 0.01$. (b) Interface population as a function of time in reverse bias for an initial state of $\ket{\us \us}$, $\delta = 0.1$, $\Delta = 50$, and $J_{34} = -(\Delta + 1)J$.
(c) $\mathcal{R}$ as a function of $h_n$ and $\delta'$ with 
$\delta = 0.03$ and $\Delta=5$. (d) Rectification $\mathcal{R}$ as a function of the coherence time $T$ for $\delta = 0.1$ and $\Delta = 5$. This is done without error-correction (dashed black) and with error-correction (solid red).
For plots (a), (c), and (d) the parametrization $J_{34} = J^{\textrm{c}}_{34}(\Delta)$ was used.
}
\label{figure3}
\end{figure}

To explain the above observations, we first note that the biggest change in current occurs in reverse bias. Therefore, this is the situation we will focus on. To motivate entanglement as a cause of the large rectification, we plot the entanglement measure $\mathcal{T}$ for the interface alongside the contrast $\mathcal{C}$ in Fig.~\ref{figure3}{(a)}. The entanglement measure used is called the concurrence \cite{Hill:97, PhysRevLett.80.2245}
\begin{equation} 
\mathcal{T}(\hat{\rho}_{\textrm{ss},\textrm{r}}^{(34)}) = \max (0, \lambda_1 - \lambda_2 - \lambda_3 - \lambda_4), 
\end{equation}
where $\lambda_1, ..., \lambda_4$ are eigenvalues, in decreasing order, of the non-Hermitian matrix 
\begin{equation*} 
\hat{\rho}_{\textrm{ss},\textrm{r}}^{(34)} \left(\sigma^{(3)}_y \sigma^{(4)}_y\right) \hat{\rho}_{\textrm{ss},\textrm{r}}^{(34)*} \left(\sigma^{(3)}_y \sigma^{(4)}_y\right).
\end{equation*}
The concurrence is a widely used measure of entanglement which is $1$ only for a maximally entangled state. The state of the interface $\hat{\rho}_{\text{ss,r}}^{(3,4)} = \tr_{(1,2,5,6)} \{\hat{\rho}_{\text{ss,r}}\}$ is found by tracing over the Hilbert space of spins 1, 2, 5, and 6. In Fig.~\ref{figure3}{(a)}, we observe a strong correlation between the amount of entanglement and the diode being in a working regime. An inspection of the density matrix shows that the entanglement is in the form of the entangled Bell state $\ket{\Psi_-}~=~\left(\ket{\us \ds}-\ket{\ds \us}\right)/\sqrt{2}$. This is further backed by the steady-state population for the interface $P_{\text{r}}(\ket{\Psi_-})~=~\langle \Psi_- | \hat{\rho}_{\text{ss,r}}^{(3,4)}| \Psi_-\rangle$ plotted in Fig.~\ref{figure3}{(a)}. 
To see that the rectification is indeed due to entanglement through $\ket{\Psi_-}$, the explanation can be broken into two parts:
\begin{enumerate}
\item The entangled state $\ket{\Psi_-}$ prevents transport between the baths.
\item In reverse bias, the interface is driven into the entangled state $\ket{\Psi_-}$.
\end{enumerate}
For the first part, the Hamiltonian is used on a compound state where spins 1 and 2 are down due to the cold bath, the interface is in the entangled state, and spins 5 and 6 are in a general state
\begin{equation}
\hat{H} \ket{\ds \ds \Psi_-\, S} = E \ket{\ds \ds \Psi_-\, S} + \sqrt{2} \delta J \ket{\ds \us \ds \ds S},\label{eq:interference}
\end{equation}
where $S \in \{\ket{\ds \ds}, \ket{\Psi_-}, \ket{\Psi_+}, \ket{\us \us}\}$. This state is unaffected by the left bath, and the right bath can only couple these four states to each other. Remarkably, the state $\ket{\ds \ds \Psi_- \, S}$ is close to being a stationary state of $\hat{H}$ for $\delta \ll 1$ with energy $E$. Therefore, the entangled state can not propagate to spin 2 due to destructive interference, and the transition, $\ket{\ds \ds \Psi_-\, S} \leftrightarrow \ket{\ds \us \ds \ds S}$, is further forbidden by energy conservation. Furthermore, any spin excitation at spin 5 cannot propagate to the interface due to perfect destructive interference. This destructive interference can be summed up by the relations
\begin{subequations} 
\label{eq:int_rel}
\begin{alignat}{1}
\left(\hat{X}_{23} + \hat{X}_{24} \right) \ket{\downarrow \downarrow \! \Psi_- \, S} &= 0, \\
\left(\hat{X}_{45} + \hat{X}_{35} \right) \ket{\downarrow \downarrow \! \Psi_- \, S} &= 0,
\end{alignat}
\end{subequations}
which show up directly when deriving Eq.~\eqref{eq:interference}. Therefore, a spin excitation is prevented from traveling between the baths resulting in a suppressed spin current. Due to the second part of Eq.~\eqref{eq:interference}, the entangled state will decay weakly, and therefore, we expect $\delta \ll 1$ to be preferable. 

For the second part of the explanation, we already verified that the interface is indeed driven into the entangled state $\ket{\Psi_-}$. The reason for the system being driven into the entangled state can be found by looking at the transition
\begin{equation}
\label{eq:trans}
\ket{\ds \ds \us \us S} \leftrightarrow \ket{\Psi_\pm \Psi_- \,S} \rightarrow \ket{\ds \ds \Psi_- \,S}.
\end{equation}
In Fig.~\ref{figure3}{(b)}, the population of the interface spins is plotted as a function of time for an initial state of $\ket{\ds \ds \us \us \ds \ds}$. A population of $P_r(\ket{\Psi_-}) \simeq 0.9$ is quickly reached, $t<50J^{-1}$, at timescales which are much shorter then the usual relaxation time $t \sim 10^3J^{-1}$, see Appendix \hyperref[AppA]{A}. We can write the matrix elements of the Hamiltonian as
\begin{subequations}
\begin{alignat}{1}
\langle \ds \ds \us \us S | \hat{H} | \ds \ds \us \us S\rangle &= \Delta J + E_S \\ 
\langle \ds \ds \us \us S|\hat{H}|\Psi_\pm \Psi_- \,S\rangle &= \mp \delta J, \\ \langle \Psi_\pm \Psi_- \,S|\hat{H}|\Psi_\pm \Psi_- \,S\rangle &= -\Delta J - 2J_{34} \pm 2J + E_S,
\end{alignat}
\end{subequations}
where $E_S$ is the energy of $\ket{S}$. For the first part of the transition, $\ket{\ds \ds \us \us S} \leftrightarrow \ket{\Psi_\pm \Psi_- \, S}$, to be favorable, it needs to obey energy conservation, and the transition matrix element needs to be large. The two states are at resonance when
\begin{subequations}
\begin{alignat}{1}
\langle \ds \ds \us \us S|\hat{H}|\ds \ds \us \us S\rangle &= \langle \Psi_\pm \Psi_- \,S|\hat{H}|\Psi_\pm \Psi_- \,S\rangle \\ 
 \text{or} \quad J_{34} &= -(\Delta \pm 1)J.
\end{alignat}
\end{subequations}
These two solutions correspond to the two bands of large rectification in Fig.~\ref{figure2}{(a)} mentioned previously. The minus solution is almost the same as $J_{34}^c(\Delta)$ with a discrepancy of $0.3$. From the matrix element $\langle \ds \ds \us \us S|\hat{H}|\Psi_\pm \Psi_- \,S\rangle$, we would expect a larger $\delta$ to yield a larger rectification. However, picking $\delta$ is clearly a balance. While a larger $\delta$ results in the state $\ket{\Psi_-}$ recovering faster, a larger $\delta$ also results in a decay of $\ket{\Psi_-}$ as can be seen in Eq.~\eqref{eq:interference}. Apparently, the present setup requires the smallest non-zero value of $\delta$ achievable for this balance to be optimal i.e. $\delta \rightarrow 0$. If decoherence is included, this balance is changed and a larger $\delta$ is required to compensate. The same is true for other imperfections e.g. a magnetic field on spin 3 or 4. Now that the mechanism is understood, many alternative versions and an expansions of the setup can be found using the same logic. Some of these are studied in the Appendices \hyperref[AppB]{B}-\hyperref[AppC]{C}.

\subsection{Sensitivity to perturbations and noise}

\noindent Next, we study the sensitivity of the rectification to local magnetic fields, coupling strength perturbations, and finite coherence times. 
Since the rectification mechanism relies on entanglement and interference, we can expect the rectification to be sensitive towards decoherence and pertubations that break the interference conditions in Eq.~\eqref{eq:int_rel}. Therefore, we can expect spins 3 and 4 to be most sensitive to magnetic fields, 
while the coupling of spins 4 and 5 should be the more sensitive coupling
parameter.
Hence, we add to Eq.~\eqref{hamiltonian_main} perturbations of the form
$\hat{H}' = \sum_{n = 1}^6 h_n \hat{\sigma}_z^{(n)} + \delta' J \hat{X}_{45}$.
Fig.~\ref{figure3}{(c)} shows $\mathcal{R}$ as a function of $h_n$ and $\delta'$, where for each line, the other perturbations are kept zero.
As expected, the rectification is stable towards changes in $h_1$, $h_2$, $h_5$, and $h_6$. The largest $\mathcal{R}$ requires magnetic fields $h_3$ and $h_4$ of 
less than $20\%$ of $J$, which is within experimental 
precision for, e.g., superconducting circuits \cite{PhysRevLett.113.220502}. 
Fig.~\ref{figure3}{(c)} also shows $\mathcal{R}$ as a function of $\delta'$ and indicates that $\delta' < \delta$ is the region of 
large rectification. The rapid decrease in $\mathcal{R}$ could 
be used to detect variations in couplings in the system. The sensitivity towards variation in $\gamma$ and the special case of $\delta = \delta'$ is studied in Appendices \hyperref[AppD]{D}-\hyperref[AppE]{E}.
Decoherence is included by adding to Eq.~\eqref{Lindblad} the perturbation 
\begin{equation} 
\begin{aligned}
\mathcal{L}'[\hat{\rho}] &= \frac{1}{T} \sum_{n = 1}^6 \left( \hat{\sigma}_-^{(n)} \hat{\rho} \hat{\sigma}_+^{(n)}  - \frac{1}{2} \left\{ \hat{\sigma}_+^{(n)}\hat{\sigma}_-^{(n)} , \hat{\rho} \right\} \right) \\
&\hspace{1.44cm} + \frac{1}{4T} \sum_{n = 1}^6 \left( \hat{\sigma}_z^{(n)} \hat{\rho} \hat{\sigma}_z^{(n)}  - \frac{1}{2} \left\{ \hat{\sigma}_z^{(n)}\hat{\sigma}_z^{(n)} , \hat{\rho} \right\} \right).
\end{aligned}
\end{equation}
The coherence time is $T = T_1 = T_2$ for both decay $T_1$ and dephasing $T_2$.
In Fig.~\ref{figure3}{(d)}, the rectification for the diode is plotted as a function of $T$. Current quantum technologies have an estimated $T J \sim 4\cdot 10^4$ for superconducting circuits \cite{devoret2013} and $T J \sim 10^4$ for trapped ions \cite{HAFFNER2008155, Johanning_2009}, for which the entanglement enhanced diode performs better than the linear version (see the Appendix \hyperref[AppF]{F} for more detail). For 
near-term devices, we provide an autonomous error correction scheme that may enhance the rectification as seen in Fig.~\ref{figure3}{(d)}, the details of which are given in Appendix \hyperref[AppF]{F}. With improving coherence times in future devices, the rectification and thus the benefit of the diode increases essentially linearly.

\subsection{Generalization to heat currents}

\begin{figure}[]
\begin{center}
\includegraphics[width=1 \linewidth, angle=0]{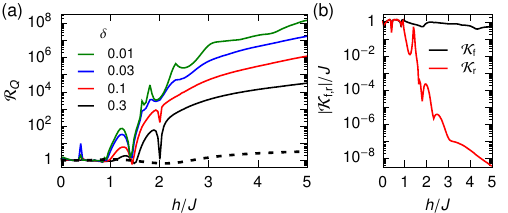}
\end{center}

\caption{(a) Heat current rectification as a function of $h$ for different values of $\delta$ (solid lines). The dashed line displays $\mathcal{R}_Q$ for a linear chain with spin 3 removed. (b) Steady-state heat currents for $\delta = 0.01$. For both plots $J_{34} = h + 1.3J$ was used. The keys obey the vertical ordering of the graphs.
}
\label{figure4}
\end{figure}

\noindent Finally, we generalize to heat currents. The Hamiltonian is modified such that we still have the energy gap created by the Z-coupling but break the spin flip symmetry
\begin{equation}
\label{hamiltonianQ_main}
\hat{H}_Q = \hat{H}(\Delta = 0) + h \left( \hat{\sigma}_z^{(1)} + \hat{\sigma}_z^{(2)} \right) + \Omega \sum_{i = 1}^6 \hat{\sigma}_z^{(i)}.
\end{equation}
To study rectification of heat currents in the system, we couple the system to thermal baths at finite temperature. One can define the heat current as the heat exchanged between the system and one of the baths, 
\begin{equation}
\mathcal{K} = \mathrm{tr}\left\{\hat{H}_{Q} \mathcal{D}_1[\hat{\rho}_{ss}]\right\} =- \mathrm{tr}\left\{ \hat{H}_{Q} \mathcal{D}_6[ \hat{\rho}_{ss}]\right\}.
\end{equation}
Like before the heat rectification is defined as $\mathcal{R}_Q = - \mathcal{K}_\mathrm{f}/\mathcal{K}_\mathrm{r}$. The system is now coupled to thermal baths addressing the eigenstates of the entire system instead of just those of spin 1 and 6
\begin{subequations}
\begin{alignat}{1}
\mathcal{D}_n[\hat{\rho}] &= {\textstyle \frac{1}{2}} \hspace{-0.35cm} \sum_{\omega, \omega'}^{|\omega -\omega'| \ngg \tau_R^{-1}} \hspace{-0.35cm} \gamma_n (\omega) \left( \hat{A}_n(\omega) \hat{\rho} \hat{A}_n^\dagger(\omega') - \hat{A}_n^\dagger(\omega')\hat{A}_n(\omega) \hat{\rho} \right) + h.c.  \\
\hat{A}_n(\omega) &= \sum_{\omega=\varepsilon'-\varepsilon} \Pi(\varepsilon) \sigma_x^{(n)} \Pi(\varepsilon')
\end{alignat}
\end{subequations}
for $n\in \{1,6\}$. $\Pi(\varepsilon)$ is the projection operator onto the space of eigen states of $\hat{H}_{Q}$ with eigen energy $\varepsilon$. The first sum is done over all pairs of frequencies for which $|\omega - \omega'|$ is not much greater than the inverse relaxation time of the diode $\tau_R^{-1}$. The second sum is carried out over all pairs of projection operators $\Pi (\varepsilon)$ and $\Pi (\varepsilon')$ with the energy difference $\omega = \varepsilon' - \varepsilon$.
The coupling strength for transitions of frequency $\omega$ is
\begin{equation} 
\gamma_n(\omega) = \left\{ \begin{matrix}
J(\omega) \left(1 + N_n(\omega)\right) & \omega \geq 0\\
J(\omega) N_n(\omega) & \omega < 0
\end{matrix} \right. .
\end{equation}
$N_n(\omega) = \left( \exp(|\omega|/T_n) -1 \right)^{-1}$ is the Bose-Einstein distribution describing the mean number of phonons in the bath mode with frequency $\omega$, and $J(\omega)$ is the spectral function. Here we consider an ohmic bath for which $J(\omega) = \gamma |\omega|$.
This is called the global master equation because the baths address eigen states of the entire system. For $\Omega \gg h, J_{34}, J$, the baths are approximately local similar to the original model, and the rectification values are similar to those seen in Fig.~\ref{figure2}{(b)}. For $\Omega = 0$, spins 1 and 6 are coupled to two thermal baths using the global master equation.
For $\Delta < 5$, the optimal rectification is achieved for $J_{34} = h + 1.3J$. The rectification of the heat diode is shown in Fig.~\ref{figure4}{(a)} for a cold bath of temperature $0.1J$ and a hot bath of temperature $10.1J$. For comparison, the rectification of the reduced system where spin 3 is removed is shown with a dashed line in Fig.~\ref{figure4}{(a)}. In Fig.~\ref{figure4}{(b)}, it is verified that large rectification is due to a suppression of $\mathcal{K}_r$. The proposed diode thus generalizes very well to heat currents where rectifications of $>10^8$ can be reached. For a more detailed analysis of the global master equation approach and the parametrization, $J_{34} = h + 1.3J$, see the Appendix \hyperref[AppG]{G}.

\section{Conclusions}

We have proposed a new class of rectifier explointing the quantum mechanical effects of entanglement and interference. 
The rectifier is comprised of two segments of quantum spin chain coupled through a two-way interface. 
Rectification factors of $\mathcal{R} > 10^5$ for realistic anisotropy factors $\Delta < 5$ was achieved. The mechanism was found to rely on an entangled Bell state developing in reverse bias thus blocking transport. 
The effect was found for a large set of parameters and in noisy environments, and the rectification is present even for heat currents using the global master equation approach.
The entanglement-enhanced rectification diode proposed here has many variations (see Appendix \hyperref[AppB]{B}) and generalizes to larger systems of $N=7$ spins (see Appendix \hyperref[AppC]{C}). 
It is built within a generic model with no particular implementation in mind and could be realized with several of the current quantum technology platforms including surface chains of atoms \cite{hirjibehedin2006,doi:10.1126/science.1201725}, trapped ions \cite{PhysRevLett.92.207901,blatt2012}, semiconductor structures, doped silicon systems, quantum dots, and NV centers \cite{Morton2011,Awschalom1174}, Rydberg atoms \cite{saffman2010}, and superconducting circuits \cite{devoret2013}. 
Finally, we note that the rectification found here is due to loss in reverse-bias conductance from the increase in exchange between spin 3 and 4. 
Interestingly, this unintuitive behavior has a classical analog known as Braess paradox \cite{doi:10.1287/trsc.1050.0127} and is seen in lesser extend in traffic, mechanical, electrical, and microfluidic networks \cite{Cohen1991,Nagurney_2016,Case2019}.

\begin{acknowledgments}
We thank Philip Hofmann and Jill Miwa for feedback on the text, as well as Kristen Kaasbjerg and Antti-Pekka Jauho for discussion. We are particularly grateful to Sai Vinjanampathy, Suddhasatta Mahapatra, and Bhaskaran Muralidharan for careful feedback and discussions on the setup and technical details. K.P. and N.T.Z. acknowledge funding from The Independent Research Fund Denmark DFF-FNU. A.C.S. acknowledges financial support from the São Paulo Research Foundation (FAPESP) (Grant No 2019/22685-1 and 2021/10224-0). L.B.K. acknowledges financial support from the Carlsberg Foundation.
\end{acknowledgments}

\appendix

\renewcommand{\thesubsection}{\Alph{subsection}}

\section*{Appendix A: Uniqueness of the Steady State}
\label{AppA} 

\begin{figure}[t]
	\centering

	\includegraphics[width=1 \linewidth, angle=0]{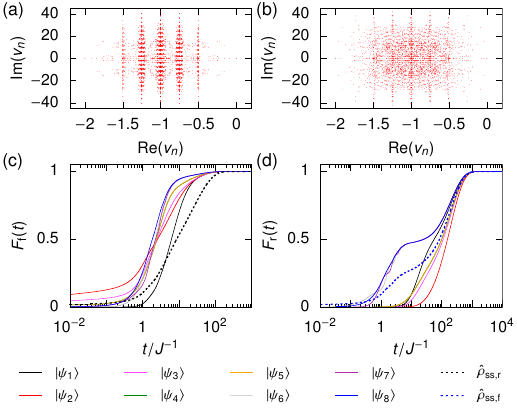}
	
	\caption{Eigenvalues $\nu_n$ for the Lindbladian $\mathbb{L}$ plotted for forward (a) and reverse bias (b). Below the fidelity $F_{\mathrm{f/r}}(t)$ between the density matrix $\hat{\rho}(t)$ and the desired steady state $\hat{\rho}_{\mathrm{ss,f/r}}$ for each of the 10 states as initial states is plotted both in forward bias (c) and reverse bias (d). For this the values $\delta = 0.1$, $\Delta = 5$, $J_{34} = J^{c}_{34}(\Delta=5)$ and $\gamma = J$ were used.}
	\label{figure5}
\end{figure}

In this section, we focus on the question: \textit{Is the steady state dependent on the initial state?} To answer this question, we define the super-operator $\mathcal{L}$, which describes the evolution of the density matrix $\hat{\rho}$ of the diode through
\begin{equation}
	\frac{\partial \hat{\rho}}{\partial t} = \mathcal{L}[\hat{\rho}] = -i [\hat{H}, \hat{\rho}] + \mathcal{D}_{1}[\hat{\rho}] + \mathcal{D}_{6}[\hat{\rho}] \label{Lindblad_app}
\end{equation}
as defined in the main text.
Likewise, the Hamiltonian is given by
\begin{equation}
\hat{H}/J = \hat{X}_{12} + (1+\delta) \hat{X}_{23} + \hat{X}_{24} + J_{34}/J \hat{X}_{34} + \hat{X}_{35} + \hat{X}_{45} + \hat{X}_{56} + \Delta \hat{Z}_{12} ,
\end{equation}
where $\hat{X}_{ij} = \hat{\sigma}_x^{(i)} \hat{\sigma}_x^{(j)} + \hat{\sigma}_y^{(i)} \hat{\sigma}_y^{(j)}$ 
is the $XX$ spin exchange operator, while 
$\hat{Z}_{ij} = \hat{\sigma}_z^{(i)} \hat{\sigma}_z^{(j)}$ is the $Z$ coupling that induces relative
energy shifts. The Pauli matrices for the $i$th spin are denoted $\hat{\sigma}_\alpha^{(i)}$ for $\alpha = x,y,z$, and
we are using units where $\hbar=k_B=1$. The exchange coupling $J$ gives the overall scale of the problem, while the 
exchange between the interface spins is $J_{34}$.
$\mathcal{D}_{n}[\hat{\rho}]$ is another super-operator describing the action of the environment on our system and is defined by
\begin{equation}
\begin{aligned}
\mathcal{D}_n [\hat{\rho}] &= \gamma \left[ \lambda_n  \left( \hat{\sigma}_+^{(n)} \hat{\rho} \hat{\sigma}_-^{(n)}  - \frac{1}{2} \left\{ \hat{\sigma}_-^{(n)}\hat{\sigma}_+^{(n)} , \hat{\rho} \right\} \right) \right. \\
& \hspace{1.15cm} \left. + (1- \lambda_n) \left( \hat{\sigma}_-^{(n)} \hat{\rho} \hat{\sigma}_+^{(n)}  - \frac{1}{2} \left\{ \hat{\sigma}_+^{(n)}\hat{\sigma}_-^{(n)} , \hat{\rho} \right\} \right) \right]. 
\end{aligned}
\end{equation}
Thus $\mathcal{L}$ contains operators acting from both left and right making Eq.~\eqref{Lindblad_app} difficult to solve in the current form. Therefore, one can define the operation $\dket{\rho} = \mathrm{vec}(\hat{\rho})$ that stacks the columns of $\hat{\rho}$ on top of each other resulting in a vector of length $D^2 = 2^{2\cdot 6}$. For example one would get
\begin{equation}
\mathrm{vec} \begin{pmatrix}
\rho_{1,1} & \rho_{1,2}\\
\rho_{2,1} & \rho_{2,2}
\end{pmatrix} = \begin{pmatrix}
\rho_{1,1} \\
\rho_{2,1} \\
\rho_{1,2} \\ 
\rho_{2,2}
\end{pmatrix}. 
\end{equation}
Using this operation one can show that
\begin{equation}
\mathrm{vec}(\hat{A} \hat{\rho} \hat{C}) = (\hat{C}^\dag \otimes \hat{A}) \mathrm{vec}(\hat{\rho}) .
\end{equation}
With this identity we can write Eq. \eqref{Lindblad_app} as
\begin{equation}
\label{eq:linlind}
\frac{\partial}{\partial t} \dket{\rho} = \hat{\mathbb{L}} \dket{\rho},
\end{equation}
where $\hat{\mathbb{L}}$ is now a $D^2 \times D^2$ matrix, with $D=2^6$, that acts on $\dket{\rho}$ only from the left. It can be written as
\begin{subequations}
\begin{alignat}{1}
\hat{\mathbb{L}} &= -i \left( 1\!\!1 \otimes \hat{H} - \hat{H} \otimes 1\!\!1 \right) + \hat{\mathbb{D}}_1 + \hat{\mathbb{D}}_6\\ 
\hat{\mathbb{D}}_{n} &= \gamma \left[\lambda_n \left( \hat{\sigma}^+_{n} \otimes \hat{\sigma}^+_{n}  - \frac{1}{2} \left(1\!\!1 \otimes \hat{\sigma}^-_{n} \hat{\sigma}^+_{n} + \hat{\sigma}^-_{n} \hat{\sigma}^+_{n} \otimes 1\!\!1\right) \right) + \right. \nonumber \\  &\hspace{0.8cm}  \left. (1-\lambda_n) \left( \hat{\sigma}^-_{n} \otimes \hat{\sigma}^-_{n}  - \frac{1}{2} \left( 1\!\!1 \otimes \hat{\sigma}^+_{n} \hat{\sigma}^-_{n} + \hat{\sigma}^+_{n} \hat{\sigma}^-_{n} \otimes 1\!\!1 \right) \right) \right] .
\end{alignat}
\end{subequations}
Since $\mathbb{L}$ is not Hermitian it is not necessarily diagonalizable, but if we assume that it is, we can write the initial state as an expansion in right eigen-vectors $\dket{e_n}$ of $\mathbb{L}$
\begin{equation}
\dket{\rho (0)} = \sum_{n=1}^{D^2} c_n \dket{e_n}. 
\end{equation}
Using this the differential equation \eqref{eq:linlind} can easily be solved
\begin{equation} 
\dket{\rho (t)} = \sum_{n=1}^{D^2} c_n e^{\nu_n t} \dket{e_n}.
\end{equation}
Here $\nu_n$ are eigenvalues of $\mathbb{L}$ such that $\mathbb{L} \dket{e_n} = \nu_n \dket{e_n}$. The eigenvalues $\nu_n$ are generally complex. The imaginary part of $\nu_n$ thus gives a time-dependent phase. The real part of $\nu_n$ ($\mathrm{Re}(\nu_n) \leq 0$) gives an exponential decay of the corresponding eigen-vector until after sufficient time only eigen states with eigenvalue $\nu_n = 0$ are left. Therefore, the steady state is found to be a zero eigen-vector of $\mathbb{L}$
\begin{eqnarray}
	\mathbb{L} \dket{\rho_{\text{ss}}} = 0.
	\label{eig_eq}
\end{eqnarray}
If only one such vector exists, all initial states will eventually decay to this vector and consequently the steady state will be unique. If more than one null-vector exists, the steady state will depend on the specific expansion coefficients $c_n$ of the initial state. Usually only one null-vector exists \cite{2018arXiv180200010A}, but this is not a given. For $\delta = 0$, the system has up-down symmetry resulting in multiple steady states. As an illustration we have plotted the $D^2 = 2^{2\cdot 6
}$ eigenvalues of $\mathbb{L}$ in forward and reverse bias in Figs. \ref{figure5}{(a)-(b)} respectively. Even though it is difficult to see from the figure, it is easily verified from the data used to plot Figs. \ref{figure5}{(a)-(b)} that there is in fact only one null eigen-vector and thus only one unique steady state. The discussion here assumes that $\mathbb{L}$ can be diagonalized. A more rigorous approach can be used in general \cite{Alicki:Book,PhysRevA.71.012331,PhysRevA.94.042131,PhysRevA.99.062320}, but the above is sufficient for the present problem.

To further emphasize that the system discussed here does in fact only exhibit one steady state for $\delta \neq 0$, we consider some different initial states. The initial states considered are
\begin{subequations}
\label{States}
\begin{alignat}{1}
\ket{\psi_{1}} &= \ket{\uparrow \uparrow \uparrow \uparrow \uparrow \uparrow} \text{ , }
\quad \ket{\psi_{2}} = \ket{\downarrow \downarrow\downarrow \downarrow\downarrow \downarrow} \\
\ket{\psi_{3}} &= (\ket{\psi_{1}}+\ket{\psi_{2}})/\sqrt{2}
 \text{ , }\quad \ket{\psi_{4}} = \ket{++++++} \text{ , } \\
\ket{\psi_{5}} &= \ket{- - - - - -} 
 \text{ , }\quad \ket{\psi_{6}} = \ket{\uparrow \downarrow \uparrow \downarrow \uparrow \downarrow} \text{ , } \\
\ket{\psi_{7}} &= \ket{\downarrow \uparrow \downarrow \uparrow \downarrow \uparrow} 
 \text{ , }\quad \ket{\psi_{8}} = (\ket{\psi_{6}}+\ket{\psi_{7}}) /\sqrt{2} \text{ , }
\end{alignat}
\end{subequations} 
where $\ket{\pm} = (\ket{\uparrow} \pm \ket{\downarrow})/\sqrt{2}$ and the state $\ket{\psi_{3}}$ is the maximally entangled GHZ (Greenberger--Horne--Zeilinger) state for 6 spins. Furthermore, we also consider starting from the steady state in forward bias $\hat{\rho}_{\mathrm{ss},\mathrm{f}}$ and reverse bias $\hat{\rho}_{\mathrm{ss, r}}$. First, we compute the steady state $\hat{\rho}_{\text{ss}}$ by solving the eigenvalue problem from Eq.~\eqref{eig_eq}. From this result, we numerically evolve each state in Eq.~\eqref{States} in time to obtain the density operator at a later time $\hat{\rho}(t)$. Then we compute the distance measure from each state $\hat{\rho}(t)$ to $\hat{\rho}_{\text{ss}}$ using the fidelity measure as provided by
\begin{equation} 
F_{\mathrm{f/r}}(t) = \left(\mathrm{tr} \sqrt{\sqrt{\hat{\rho}_{\text{ss,f/r}}} \hat{\rho}(t) \sqrt{\hat{\rho}_{\text{ss,f/r}}}} \right)^2.
\end{equation}
The results in Figs.~\ref{figure5}{(c)-(d)} show that the steady state for each of the initial states $\hat{\rho}_{n,\text{ss}}$ in Eq.~\eqref{States} is close to $\hat{\rho}_{\text{ss}}$ after sufficient time.

\section*{Appendix B: Alternative Versions of the Diode}
\label{AppB}

\begin{figure}[t]
\centering
\includegraphics[width=1 \linewidth, angle=0]{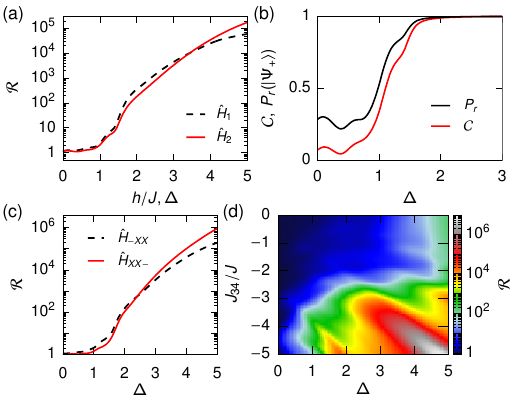}
\caption{(a) $\mathcal{R}$ for the two alternative versions of the diode defined by $\hat{H}_1$ and $\hat{H}_2$. For $\hat{H}_1$ we vary $h$ with $J_{34} = h + 1.3J$ while for $\hat{H}_2$ we vary $\Delta$ with $J_{34} = (\Delta + 1.3)J$. (b) Reverse bias population $P_\text{r}$ (top) and contrast $\mathcal{C}$ (bottom) as a function of $\Delta$ for the Hamiltonian $\hat{H}_2$ with the parametrization $J_{34} = (\Delta + 1.3)J$. (c) Rectification $\mathcal{R}$ as a function of $\Delta$ for the two extensions corresponding to the Hamiltonians $\hat{H}_{-XX}$ and $\hat{H}_{XX-}$, where $J_{34} = J_{34}^{c}(\Delta)$. (d) Rectification $\mathcal{R}$ as a function of $\Delta$ and $J_{34}$ for the Hamiltonian $\hat{H}_{XXZ-}$. For all plots $\delta = 0.01$ and $\gamma = J$. 
}
\label{figure6}
\end{figure}

So far, we have explored one configuration of the setup that exhibits large rectification and which is given by the Hamiltonian \eqref{hamiltonian_main}. However, many other sets of parameters will work just as well, some of which will lend themselves more suitably to different physical implementations. Here we want to explore some of these alternative versions.

So far, we have chosen $\Delta > 0$, which then led to the critical value $J_{34} = -(\Delta + 1.3)J$. Alternatively, we could have chosen $\Delta < 0$ in which case we would have gotten the critical value $J_{34} = (-\Delta + 1.3)J$. This corresponds to the transformation $(\Delta, J_{34}) \rightarrow (-\Delta, -J_{34})$. A more general parametrization would, therefore, be
\begin{equation} 
J_{34} = \left\{\begin{matrix}
(-\Delta + 1.3)J & \Delta < 0\\
-(\Delta + 1.3)J & \Delta > 0
\end{matrix} \right. .
\end{equation}
From Fig.~\ref{figure2}{(a)} it is seen that even more parametrizations give large rectification. However, for simplicity we will stop here.

The purpose of the Z-coupling between spin 1 and 2 is to create an energy gap between the state where both spins are down and the state where one spin excitation is present. This energy gab can instead be created with local magnetic fields. Thus we may define the new Hamiltonian 
\begin{equation}
\begin{aligned}
\hat{H}_1/J &= \hat{X}_{12} + (1+\delta) \hat{X}_{23} + \hat{X}_{24} + J_{34}/J \hat{X}_{34} + \hat{X}_{35} \\
&\hspace{2.3cm} + \hat{X}_{45} + \hat{X}_{56} + h/J \left(\hat{\sigma}_z^{(1)} + \hat{\sigma}_z^{(2)}\right)  .
\end{aligned}
\end{equation}
Note that $\hat{H}_1 = \hat{H}_Q$ is the Hamiltonian used for heat rectification in the main text with $\omega = 0$. With this Hamiltonian one parametrization for $h \leq 5J$ becomes 
\begin{equation} 
J_{34} = \left\{\begin{matrix}
-(- h + 1.3J) & h < 0\\
h + 1.3J & h > 0
\end{matrix} \right. .
\end{equation}
The sign difference between this and the parametrization from before is due to how a Z-coupling and a magnetic field creates the energy gap. If two spins are coupled through a Z-coupling with strength $\Delta J$, then the energy gap between the state $\ket{\downarrow \downarrow}$ and the state $\ket{\downarrow \uparrow}$ is $-2\Delta J$. If these two spin are instead coupled to a magnetic field with strength $h$, then the energy gap between the same two states is $2h$. Thus going from a Z-coupling to a local magnetic field we need to set $h = -\Delta J$. The rectification for this model is plotted in Fig. \ref{figure6}{(a)} (black dashed line), showing that this model gives rectification values of the same order of magnitude as the original model.

Another more subtle alternative version is defined by the Hamiltonian
\begin{equation} 
\hat{H}_2/J = \hat{X}_{12} - (1+\delta) \hat{X}_{23} + \hat{X}_{24} + J_{34}/J \hat{X}_{34} - \hat{X}_{35} + \hat{X}_{45} + \hat{X}_{56} + \Delta \hat{Z}_{12}  .
\end{equation}
For this Hamiltonian one possible parametrization for $\Delta \leq 5$ becomes
\begin{equation} 
J_{34} = \left\{\begin{matrix}
-(-\Delta + 1.3)J & \Delta < 0\\
(\Delta + 1.3)J & \Delta > 0
\end{matrix} \right. .
\end{equation}
To explain this we note that the state $\ket{\Psi_-}$ no longer closes the diode. Instead one can go through the same steps as in the main text to show that the state $\ket{\Psi_+}$ now causes the diode to close. Therefore, the roles of $\ket{\Psi_-}$ and $\ket{\Psi_+}$ switch around such that $\ket{\Psi_-}$ now needs to be in resonance with the rest of the diode. This is insured by letting $J_{34} \rightarrow -J_{34}$ which leads to the above parametrization. To illustrate this we have plotted $\mathcal{R}$ for this version in Fig. \ref{figure6}{(a)} (red solid line). Furthermore, the contrast $\mathcal{C}$ (see main text) and fidelity between $\hat{\rho}_+ = \op{\Psi_+}$ and the reduced steady state density matrix in reverse bias $\hat{\rho}_{\textrm{ss},\textrm{r}}^{(34)} = \mathrm{tr}_{(1, 2, 5, 6)} [\hat{\rho}_{\textrm{ss},\textrm{r}}]$ is plotted in Fig. \ref{figure6}{(b)}.

\section*{Appendix C: Scalability of the Diode}
\label{AppC}

Here we test the scalability of the diode proposed in the main text. Since open quantum systems become difficult to simulate very fast when increasing the number of spins, the strategy is as follows: Understanding the rectification mechanism for the six-spin diode, we can generalize to larger systems. Afterwards, the generalized theory is compared to results from three different seven-spin versions. The diode can be expanded at either end. Adding a spin to the right will change the energy spectrum of the right chain segment. However, this energy spectrum is unimportant for the diode mechanism, and adding additional spins to the right should neither increase nor decrease the rectification. Adding a spin to the left will change the energy spectrum of the left chain segment, which is important for the diode mechanism. For the largest rectification, the parameters of the left chain segment has to be chosen appropriately such that there is a transition resonant with the interface transition, $\ket{\us \us} \leftrightarrow \ket{\Psi_-}$. While the largest rectification is found at resonance, large rectification is achieved for a large set of parameters, see Fig.~\ref{figure2}{(a)}. Therefore, spins can be added to the left with coupling strength $\sim J$ without dramatically affecting the rectification.  Next, we look at the seven-spin version, where there are three such obvious choices. First, one could add a $7$th spin at the end of the chain with an XX coupling of strength $J$ between the $6$th and $7$th spin. Thus the new Hamiltonian is given from the original Hamiltonian $\hat{H}$ defined in Eq. \eqref{hamiltonian_main}
\begin{equation} 
\hat{H}_{-XX}/J = \hat{H}/J + \hat{X}_{67}.
\end{equation}
This new system will then be coupled to the heat baths through spin $1$ and $7$, replacing $\mathcal{D}_6[\hat{\rho}]$ with $\mathcal{D}_7[\hat{\rho}]$ in Eq. \eqref{Lindblad}. 
Alternatively, one could put a 0th spin at the beginning of the chain with an XX-coupling of strength $J$ between the 0th and 1st spin. Thus we define the Hamiltonian
\begin{equation} 
\hat{H}_{XX-}/J = \hat{H}/J + \hat{X}_{01}.
\end{equation} 
The last obvious way of extension is again with an extra 0th spin at the beginning of the chain but this time with an XXZ coupling, giving the Hamiltonian
\begin{equation} 
\hat{H}_{XXZ-}/J = \hat{H}/J + \hat{X}_{01} + \Delta \hat{Z}_{01}.
\end{equation}
The last two new systems will then be coupled to the heat baths through spin $0$ and $6$ replacing $\mathcal{D}_1[\hat{\rho}]$ with $\mathcal{D}_0[\hat{\rho}]$ in Eq. \eqref{Lindblad}. The found rectification factors for the first two versions can be seen in Fig.~\ref{figure6}{(c)}. Here we see that these two versions almost have the same rectification as the original chain (seen in Fig.~\ref{figure2}{(b)}) as expected from the discussion above. For the case of a spin added to the left of the chain coupled through an XXZ coupling (obeying the Hamiltonian $\hat{H}_{XXZ-}$), we can not expect the usual parametrization to hold. Therefore, the rectification is plotted as a function of both $\Delta$ and $J_{34}$ in Fig.~\ref{figure6}{(d)}. We see that the rectification is significantly higher than for the six spin diode. Already at $\Delta \sim 4$ (and $J_{34} \sim -4.5J$) do we achieve a rectification of $\mathcal{R} > 10^6$. Note that this model has a resonance between a spin excitation on spin $0$, $1$, and $2$ and the transition $\ket{\us \us} \leftrightarrow \ket{\Psi_-} $ for the interface if $J_{34} \sim -\Delta J$, similar to what is found in the main text. This explains why we get large $\mathcal{R}$ around $J_{34} \sim -\Delta J$.

\section*{Appendix D: $\delta = \delta'$ Symmetry}
\label{AppD}

\begin{figure}[t]
\centering
\includegraphics[width=1 \linewidth, angle=0]{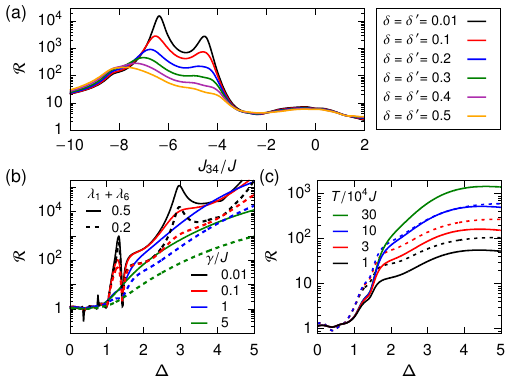}

\caption{(a) Rectification as a function of $J_{34}$ for different values of $\delta = \delta'$ and $\Delta = 5$. (b) Rectification $\mathcal{R}$ as a function of $\Delta$ for different interaction strengths $\gamma$ between the system and the bath, where $\delta = 0.01$ was used. The dashed lines denotes the case $\lambda_n \in \{0, 0.2\}$, and the solid lines denotes the case $\lambda_n \in \{0, 0.5\}$. (c) Rectification $\mathcal{R}$ as a function of $\Delta$ for different values of $T$ where $\delta = 0.1$. Solid lines denote the rectification for a model without error-correction, while dashed lines denote a model with error-correction. For plots (b)-(c) the parametrization $J_{34} = J_{34}^{c}(\Delta)$ were used. The keys obey the vertical ordering of the graphs.}
\label{figure7}
\end{figure}

In this section, we study the case where a small perturbation is added to the coupling between spin 4 and 5 such that the diode is now described by the Hamiltonian
\begin{equation} 
\begin{aligned}
\hat{H}_{\delta'}/J &= \hat{X}_{12} + (1+\delta) \hat{X}_{23} + \hat{X}_{24} + J_{34}/J \hat{X}_{34} + \hat{X}_{35} \\
& \hspace{2.9cm} + (1+\delta')\hat{X}_{45} + \hat{X}_{56} + \Delta \hat{Z}_{12}.
\end{aligned}
\end{equation}
The dependence of the rectification on $\delta'$ alone was studied in the main text, so here we will focus on $\delta = \delta'$. From the results in the main text, we expect the rectification to be lower than for $\delta'=0$. However, with $\delta = \delta'$ the system exhibits a symmetry that might be preferable to certain implementations. In Fig. \ref{figure7}{(a)}, the rectification is plotted as a function of $J_{34}$ for different values of $\delta = \delta'$. Although the rectification drops by an order of magnitude, it is still $>10^4$ for $\delta = \delta' =0.01$.

\section*{Appendix E: Interaction Strength Between Diode and Baths}
\label{AppE}
Here we study the effect of changing the interaction strength $\gamma$ between the baths and the system, as defined in Eq.~\eqref{Lindblad}, as well as the nature of the baths $\lambda_n$. This can be seen in Fig.~\ref{figure7}{(b)}, where $\mathcal{R}$ is plotted for different interaction strengths and $\lambda_n$. We see that the general behavior of the rectification is still achieved. However, for small $\gamma$, the rectification becomes more sensitive to the inner structure of the system. Generally, the rectification is increased slightly for weaker interaction strengths or larger $\lambda_1 + \lambda_6$. Interference in these types of systems are known to often disappear at stronger interaction. An example of this is molecular junctions which can be tuned such that interference effects cause the current in one bias to be zero to lowest order in the applied voltage $V$ \cite{PhysRevB.90.125413, Solomon2008}. However, to second order in $V$ the effect is broken, and rectification can be difficult to achieve \cite{C4FD00093E, Iwane2017}. In Fig.~\ref{figure7}{(b)}, we see that large rectification is achieved for couplings $\gamma$ well beyond $1J$.

\section*{Appendix F: Decoherence and Protection of the Entangled State}
\label{AppF}

To study how a limited lifetime of the spins affect rectification factors, we add both decay and dephasing on all spins. This is done by letting the density matrix evolve as
\begin{equation} 
\begin{aligned}
\frac{\partial \hat{\rho}}{\partial t} &= \mathcal{L}[\hat{\rho}] + \frac{1}{T} \sum_{n = 1}^6 \left( \hat{\sigma}_-^{(n)} \hat{\rho} \hat{\sigma}_+^{(n)}  - \frac{1}{2} \left\{ \hat{\sigma}_+^{(n)}\hat{\sigma}_-^{(n)} , \hat{\rho} \right\} \right) \\
&\hspace{1.7cm} + \frac{1}{4T} \sum_{n = 1}^6 \left( \hat{\sigma}_z^{(n)} \hat{\rho} \hat{\sigma}_z^{(n)}  - \frac{1}{2} \left\{ \hat{\sigma}_z^{(n)}\hat{\sigma}_z^{(n)} , \hat{\rho} \right\} \right),
\end{aligned}
\end{equation}
which insures that, if $\mathcal{L}[\hat{\rho}] = 0$, then the lifetime for all spins for decay ($T_1$) and dephasing ($T_2$) is $T= T_1 = T_2$. The rectification as a function of $\Delta$ is plotted in Fig.~\ref{figure7}{(c)} for different values of $TJ$. To put this plot into perspective, superconducting circuits have $T_1 \sim T_2 \sim 100\mathrm{\mu s}$ and $J/2\pi \simeq 60\mathrm{MHz}$ \cite{devoret2013} resulting in $T J \sim 4\cdot 10^4$. Ion trap based quantum computers have $T_1 \sim T_2 \sim 1\mathrm{s}$ and $J/2\pi \sim 1\mathrm{kHz}$ \cite{HAFFNER2008155, Johanning_2009} resulting in $TJ \sim 10^4$.  However, as technology improves and coherence times increase the rectification and thus the benefit of the diode also increases linearly. 

The drop in rectification is mainly due to decoherence of the entangled Bell state $\ket{\Psi_-}$. To protect against this, we can employ error-correction by forcing the transition $\ket{\ds \ds} \rightarrow \ket{\Psi_-}$. This is done by adding a shadow qubit with driving that allows the transition. We further add an excitation energy of $2\omega$ to all spins, since this will be present in most experimental setups
\begin{equation} 
\hat{H}_{EC} = \hat{H} + A \hat{\sigma}_x^{(3)} \hat{\sigma}_x^{(S)} \cos \left\{ 2( 2\omega + \Omega) t \right\} + \omega \sum_{k } \hat{\sigma}_z^{(k)},
\end{equation}
where the sum is over all spins and $\omega \gg \Omega$. Moving into the interacting picture with respect to the Hamiltonian $\hat{H}_0 = (\omega + \Omega) \hat{\sigma}_z^{(S)} + \omega \sum_{k=1}^6 \hat{\sigma}_z^{(k)}$ and performing the rotating wave approximation on terms rotating with angular frequency $\sim \omega$ we get
\begin{equation} 
\begin{aligned}
\hat{H}_{EC,I} &= e^{i\hat{H}_0 t} \left( \hat{H}_{EC} - \hat{H_0} \right) e^{-i\hat{H}_0 t} \\
&\simeq \hat{H} + A\left(\hat{\sigma}_+^{(3)} \hat{\sigma}_+^{(S)} + \hat{\sigma}_-^{(3)} \hat{\sigma}_-^{(S)} \right)  - \Omega \hat{\sigma}_z^{(S)}.
\end{aligned}
\end{equation}
Likewise the Lindblad equation and spin current operator can be transformed. We let the shadow qubit decay with rate $\gamma_S = J$ and the coupling be weak $A = 0.1J$. If we let $\Omega = -J_{34}$, the two gate spins and the shadow qubit will undergo the transition
\begin{equation} 
\ket{\ds \ds} \ket{\ds}_S \leftrightarrow \ket{\Psi_-} \ket{\us}_S \rightarrow \ket{\Psi_-} \ket{\ds}_S.
\end{equation}
Numerical simulations show that the best result is achieved for $\Omega = (\Delta + 1.2)J$. The rectification with error-correction is plotted in Fig.~\ref{figure7}{(c)} for different coherence times $T$. The error-correction works best for short coherence times where it results in about twice the rectification.

\section*{Appendix G: Heat Diode Using the Global Master Equation}
\label{AppG}

In this section, we want to explore how the diode proposed in the main text can also be used as a heat rectifier using a global master equation. First, we change the diode Hamiltonian \eqref{hamiltonian_main} slightly such that we still have the energy gap created by the ZZ-coupling but break the spin flip symmetry. This is done by using the Hamiltonian \eqref{hamiltonianQ_main}, with $\Omega = 0$, given by
\begin{equation}
\begin{aligned} 
\hat{H}_{Q}/J &= \hat{X}_{12} + (1+\delta) \hat{X}_{23} + \hat{X}_{24} + J_{34}/J \hat{X}_{34} + \hat{X}_{35}\\
&\hspace{2.2cm} + \hat{X}_{45} + \hat{X}_{56} + h/J \left( \hat{\sigma}_z^{(1)} + \hat{\sigma}_z^{(2)} \right).
\end{aligned}
\end{equation}
The energy gap is now created by a local magnetic field on spin 1 and 2 described by $h$. This Hamiltonian is the same as $\hat{H}_1$ explored in Appendix \hyperref[AppB]{B}. However, in this section we will couple spin 1 and 6 to two thermal baths and use the global master equation where the baths address the eigenstate of the total system instead of just those of spin 1 and 6. Here it does not make sense to define a spin current so instead we examine how heat is transferred through the diode. This is again done through the master equation \cite{breuer2002theory}
\begin{equation*} 
\frac{\partial \hat{\rho}}{\partial t} = -i [\hat{H}_{Q}, \hat{\rho}] + \mathcal{D}_1 [\hat{\rho}] + \mathcal{D}_6 [\hat{\rho}],
\end{equation*}
where the dissipators are now defined as
\begin{equation} 
\mathcal{D}_n[\hat{\rho}] = {\textstyle \frac{1}{2}} \hspace{-0.35cm} \sum_{\omega, \omega'}^{|\omega -\omega'| \ngg \tau_R^{-1}} \hspace{-0.35cm} \gamma_n (\omega) \left( \hat{A}_n(\omega) \hat{\rho} \hat{A}_n^\dagger(\omega') - \hat{A}_n^\dagger(\omega')\hat{A}_n(\omega) \hat{\rho} \right) + h.c.
\end{equation}
for $n\in \{1,6\}$. Here the sum is done over all pairs of frequencies for which $|\omega - \omega'|$ is not much greater than the inverse relaxation time of the diode $\tau_R^{-1}$. This is due to the secular approximation which essentially comes from a rotating wave approximation. The operators $\hat{A}_n(\omega)$ are eigen-operators of $\hat{H}_{Q}$ defined as
\begin{equation}
\hat{A}_n(\omega) = \sum_{\omega=\varepsilon'-\varepsilon} \Pi(\varepsilon) \sigma_x^{(n)} \Pi(\varepsilon')
\end{equation}
with $\Pi(\varepsilon)$ being the projection operator onto the space of eigen states of $\hat{H}_{Q}$ with eigen energy $\varepsilon$. This sum is carried out over all pairs of projection operators $\Pi (\varepsilon)$ and $\Pi (\varepsilon')$ with the energy difference $\omega = \varepsilon' - \varepsilon$. These operators describe the transitions induced by the baths with coupling strength 
\begin{equation} 
\gamma_n(\omega) = \left\{ \begin{matrix}
J(\omega) \left(1 + N_n(\omega)\right) & \omega \geq 0\\
J(\omega) N_n(\omega) & \omega < 0
\end{matrix} \right. .
\end{equation}
$N_n(\omega) = \left( \exp(|\omega|/T_n) -1 \right)^{-1}$ is the Bose-Einstein distribution describing the mean number of phonons in the bath mode with frequency $\omega$, and $J(\omega)$ is the spectral function. Here we consider an ohmic bath for which $J(\omega) = \gamma |\omega|$. We let the cold bath have temperature $T_C$ and the hot bath have temperature $T_H$. Like before, we denote $T_H = T_1 > T_6 = T_C$ as forward bias and $T_C = T_1 < T_6 = T_H$ as reverse bias.
\begin{figure}[t]
\centering
\includegraphics[width=1 \linewidth, angle=0]{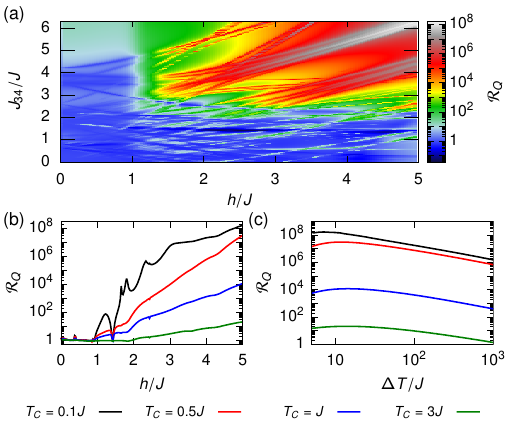}
\caption{(a) $\mathcal{R}_Q$ as a function of $h$ and $J_{34}$ for $T_C = 0.1J$ and $T_H = 10.1J$.
(b)-(c) The rectification $\mathcal{R}_Q$ plotted for different cold bath temperatures $T_C$ with hot bath temperature $T_H = T_C + \Delta T$. First $h$ is varied keeping $\Delta T = 10J$, (b), and next $\Delta T$ is varied keeping $h=5J$, (c). In (b) and (c), the values $0.1$ (top), $0.5$ (middle-top), $1$ (middle-bottom), and $3$ (bottom) was used for $T_C/J$. For all plots $J_{34} = J_{34}^Q (h)$, $\gamma = J$, and $\delta = 0.01$.
}
\label{figure8}
\end{figure}
The total change in mean energy of the diode is given by
\begin{equation} 
\frac{dE}{dt} = \frac{d}{dt} \mathrm{tr}\{\hat{H}_{Q} \hat{\rho}\} = \left\langle \frac{d H_{Q}}{dt} \right\rangle + \mathrm{tr} \left\{\hat{H}_{Q} \frac{d \hat{\rho}}{dt}\right\}.
\end{equation}
The first part is interpreted as the work done on the diode. However, since we have a constant Hamiltonian this is zero. The second part is interpreted  as the total heat going into the system. In steady state $\dot{\hat{\rho}}_{ss} = 0$, and therefore, the total heat exchanged between the diode and baths is zero. However, by noting that
\begin{equation} 
\begin{aligned}
0 &= \mathrm{tr} \left\{\hat{H}_{Q} \frac{d \hat{\rho}_{ss}}{dt}\right\} \\
&= \mathrm{tr}\left\{H_{Q} \mathcal{D}_1[\hat{\rho}_{ss}]\right\} + \mathrm{tr}\left\{ H_{Q} \mathcal{D}_6[ \hat{\rho}_{ss}]\right\},
\end{aligned}
\end{equation}
we can define the heat current as the heat exchanged between the diode and the leftbath
\begin{equation} 
\mathcal{K} = \mathrm{tr}\left\{\hat{H}_{Q} \mathcal{D}_1[\hat{\rho}_{ss}]\right\} =- \mathrm{tr}\left\{ \hat{H}_{Q} \mathcal{D}_6[ \hat{\rho}_{ss}]\right\}.
\end{equation}
Again here we denote the heat current in forward bias $\mathcal{K}_f$ and the heat current in reveres bias $\mathcal{K}_r$. Likewise, we define the rectification as
\begin{equation} 
\mathcal{R}_Q = - \frac{\mathcal{K}_f}{\mathcal{K}_r}.
\end{equation}
The heat current rectification of the diode is shown in Fig.~\ref{figure8} as a function of the relevant parameters. The contour plot in Fig.~\ref{figure8}{(a)} shows $\mathcal{R}_Q$ for a small vertical symmetry breaking of $\delta = 0.01$ as a function of $J_{34}$ and $h$. Unlike for the spin current case, we clearly see multiple resonances that makes the plot chaotic for small $J_{34}$ and $h$. However, in the upper right corner many of the resonance merge and create thicker more stable lines of large rectification of $> 10^8$. We note that the region of largest $\mathcal{R}_Q$ follows the same parametrization as before which we call $J^Q_{34}(h) = h + 1.3J$. 
In Figs.~\ref{figure8}{(b)-(c)}, the rectification as a function of the bath parameters is studied. In Fig.~\ref{figure8}{(b)}, it can be seen that largest rectification is achieved for $T_C < J$. Since $J$ sets the energy scale of the diode, for $T_C < J$ the cold bath will predominantly induce decay while the hot bath will induce both decay and excitation in the energy levels. Therefore, we expect a better diode for smaller $T_C$. In Fig.~\ref{figure8}{(c)} we see that the rectification is stable over the first order of magnitude in $\Delta T$ but decreases slightly for very large $\Delta T$.

\bibliography{bibliography}

\end{document}